# NOVEL MCP-BASED ELECTRON SOURCE STUDIES*

M. Haughey, School of Physics and Astronomy, University of Edinburgh, Edinburgh, UK

V. Shiltsev[†], G. Stancari, Fermilab, Batavia IL 60510, USA

*Abstract*

Microchannel plates (MCPs) were recently proposed as novel type of cathodes for electron guns [1], suitable for applications in design of electron lenses. We report results of the first systematic study of microchannel plate based photomultiplier time response and maximum current density tests using different sources of light pulses. The Burle 85011-501 MCP-PMT is found to have good time response properties being capable of producing nanosecond long pulses with modest maximum current density and performance strongly dependent on magnetic field strength.

## INTRODUCTION

The concept of the MCP-based electron gun – see Fig.1 - has sparkled interest because of the opportunity of effective generation and fast modulation of the electron current as well as the promise of easy transverse shaping of the beam by proper exposure of the photocathode surface by either masking of the light or by several laser beams [1, 2]. Application of such guns in the electron lenses for particle accelerators [3], including the IOTA ring at Fermilab [4], requires peak electron current densities $O(1\ \text{A/cm}^2)$ in short (~ several ns long) pulses with relatively low duty factors [5, 6].

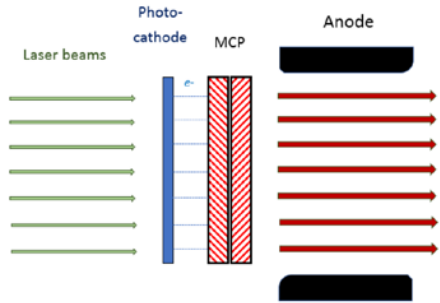

Figure 1: Layout of the MCP-based electron gun.

To use an MCP as the cathode of an electron gun would bring several marked benefits over previous thermionic cathode designs: a) the ability of MCP devices to operate at room temperature allow for improvements to the mechanical stability of the electron gun system, improved vacuum conditions and reduced beam emittance; b) MCP devices can have a quantum efficiency of around 15–30%, which is a tenfold improvement on other electron gun designs that use photosensitive materials, such as photoinjectors; c) a superior pulse width – orders of magnitudes shorter than ~100 ns of the traditional anode-modulated guns for electron lenses [3]. Here we report first measurements of the high peak electron current generation in the MCP-PMTs and effects of the magnetic field on the cathode.

## MCP-PMT TESTS

The MCP-PMT device used throughout this study is the Burle 85011-501 MCP-PMT [7]. It is a two-stage device, using two MCPs in the electron-multiplying stage to provide a typical gain value of $7\times10^5$ at an applied voltage of -2.5 kV. This MCP-PMT has 8×8=64 channels, or discrete anode pixels with 35mm$^2$ area each, which can be individually read out as a voltage signal. It is capable of a time response of around 300 ps, and the bialkali photocathode is sensitive to wavelengths in the visible region from 185 to 660 nm, with a peak sensitivity at 400 nm. The device was tested with an LED and a laser.

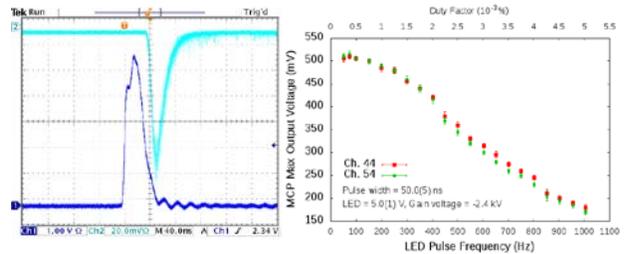

Figure 2: a) left - blue trace is signal from the function generator driving blue LED, cyan trace is output signal from one of the MCP-PMT channels (40 ns/div, 300 MHz bandwidth); b) right - MCP output pulse amplitude versus duty factor for MCP-PMT channels 44 (red) and 54 (green).

A blue LED pulsed by a function generator illuminated the Burle MCP-PMT mounted inside a dark box for about 50 ns and output signals were read out on a 300MHz bandwidth Tektronix oscilloscope, also shown. Fig. 2 a) shows a typical scope readout from these tests. The MCP-PMT output amplitude appears to have a non-linear dependence on the applied voltage and to maximize it we used the maximum manufacturer recommended value of about -2.5 kV. From the data in Fig. 2 a), the maximum output amplitude attainable is around 650 mV per pixel, that corresponds to electron current density of 650/(50 ohm)/35mm$^2$=0.04 A/cm$^2$. This value seems to be limited by the irradiance and pulse width of the LED. The dependence of MCP-PMT output amplitude on duty factor is explored by varying the frequency of LED pulses, keeping the pulse width constant. A relatively small range of frequency values was studied, due to limitations imposed by the function generator, yet a strong dependence on the duty factor is observed and is demonstrated for two channels of the MCP-PMT in Fig. 2b. Even at a frequency value of 1 kHz, corresponding to a duty factor value of $5 \times 10^{-3}$ % for pulse width of 50 ns, the MCP-PMT signal is heavily suppressed in comparison the signal at 100 Hz. The reasons for this are unknown, however on page 285 in Ref. [8] the author describes briefly that when MCP-PMT devices are operated at high voltages, the resulting gain may be large enough that the charge at the back face of the plate (at the channel exits) limits the total charge per pulse. The observations from the duty factor experiments in this study may be a demonstration of this effect: if the charge at the back plate of the MCP cannot dissipate quickly enough then as the duty factor increases there may be a build up of space charge, limiting the total output charge of the pulses. It will be crucial to consider this effect when assessing the suitability of MCP technology for the electron gun design.

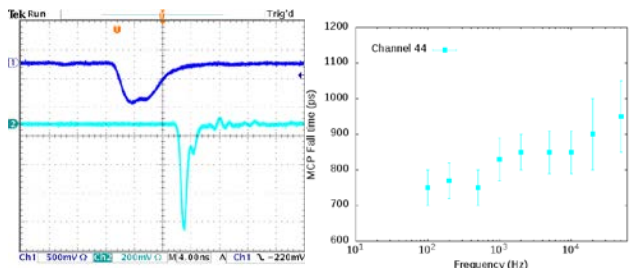

Figure 3: a) left - typical scope readout for laser test. Dark blue trace is the laser pulse signal from the laser controller. Cyan trace is output signal from MCP-PMT, channel 44 (4 ns/div, 500 MHz, 5 GS/s scope); b) right – MCP response fall time versus laser pulse frequency.

The same Burle MCP-PMT was then tested with a 635 nm pulsed laser. This laser is much faster than the LED system, and as such the shortest achievable pulse from the MCP-PMT is about 5 ns (similar to the IOTA bunch length, 3 ns). Fig. 3a) shows a typical scope readout from the laser tests, demonstrating such a pulse from the MCP-PMT. The relationship between the MCP-PMT response time with frequency is presented in Fig.3b) which indicates roughly logarithmic increase of the response time with frequency. It should be noted that the data corresponding to larger frequency values are less reliable. The reason for this is that at these values of frequency, the duty factor is large enough to suppress the MCP-PMT response, reducing the resolution of the signal on the scope. For this reason, its response time at frequencies relevant to the IOTA ring (~8 MHz) are not explored.

Despite the pulse width of the laser being an order of magnitude smaller than those produced by the LED, the maximum voltage output from the MCP-PMT was comparable to that from the LED tests - just under 400 mV as can be seen from Fig. 3a). Brief tests using a laser with a greater peak power show that this value can be increased (magnetic field tests, addressed in a later section, are conducted using a laser with a marginally higher peak power and lead to peak current densities in the region of 60 mA/cm$^2$). This suggests that an upper limit on the current density available from this particular MCP-PMT device has not yet been established, and more powerful lasers, e.g., the FAST photoinjector laser, should be tried in future tests. There might be an upper limit on the total charge available per pulse in microchannel plate-based devices (between $10^6$ and $10^7$ electrons per channel) as suggested by Ref. [8], p. 285.

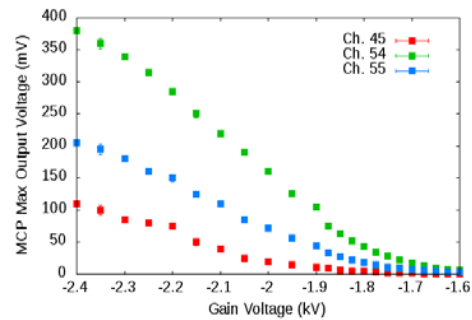

Figure 4: MCP-PMT output amplitude versus applied gain voltage. Three colours of data points correspond to different MCP-PMT channels.

Results of the studies of the effects of applied gain voltage and the laser pulse duty factor are shown in Fig.4. A linear relationship between MCP-PMT output amplitude and applied MCP-PMT voltage is observed between -2.4 kV and -2.0 kV. A far greater range of frequencies, and therefore duty factors, can be explored with the laser. The frequency of the laser is varied from 100 Hz to 1 MHz, spanning four orders of magnitude in the duty factor – see Fig.5. The data from these tests support those from the LED tests, showing increases in the duty factor to have a strong negative effect on the peak output voltage from the MCP-PMT. More is revealed about the relationship between the output voltage and the duty factor with these tests, and it is observed that the output voltage increases exponentially with decreasing duty factor over the range from 100 Hz to 1 MHz.

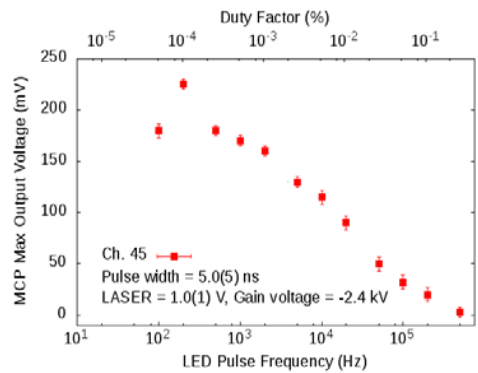

Figure 5: MCP-PMT output amplitude versus duty factor. Logarithmic scale in the horizontal axis.

The electron gun in an electron lens must operate in a solenoidal magnetic field, and so we also studied the effect of the field on the performance of MCP-PMT devices. A permanent magnet, with maximum fields strength of ∼ 700 G is used (that was comparable, though lower than typical 2 kG to 4 kG magnetic field in electron lenses). The aperture of the magnet was not large enough to accommodate the MCP-PMT and laser, and so the experiments are conducted using the fringes of the field. The magnetic field strength in the volume occupied by the MCP-PMT is controlled by varying the distance of the device from the magnet aperture in both cases. Two configurations are explored – with magnetic field perpendicular to the MCP channels, and with magnetic field and MCP channels parallel. The latter is of particular interest, as this recreates more accurately the configuration of the electron lens. Measurements of MCP-PMT output amplitude versus magnetic field strength are presented in Fig. 6, showing data corresponding to both configurations. A strong dependency is observed in both scenarios; with the MCP-PMT device performing poorly even in a relatively weak magnetic field. One MCP-PMT channel is studied in the perpendicular configuration, the results for which show a linear decay in MPC-PMT output amplitude with magnetic field strength. This particular result is in agreement with Ref. [9] where it also was found a similar stark decrease in performance quality of the Burle 85011-501 MCP-PMT in magnetic fields.

For the parallel configuration, two MCP-PMT channels are studied. The data from the parallel configuration, despite only a few points, also hint at a negative impact on the MCP-PMT performance with increasing magnetic field strength. This is likely due to the electrons being accelerated by the magnetic field back to the walls of the channel before they have sufficient energy to cause more electron emission.

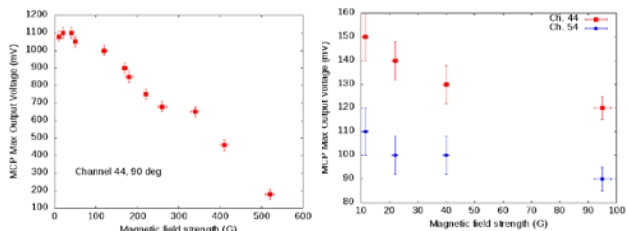

Figure 6: MCP-PMT output amplitude versus magnetic field strength for: (left) perpendicular configuration; (right) parallel configuration.

## DISCUSSION

A systematic study of a microchannel plate based photomultiplier device Burle 85011-501 MCP-PMT has been carried out using flashing LED and pulsed laser. We obtain short output electron pulses with width of few ns and a current density up to 0.04-0.06 A/cm$^2$. We find a strong dependence of the output amplitude on the LED/laser pulse duty factor and on the presence and orientation of an external magnetic field.

The goal of obtaining a current density of ∼1 A/cm$^2$ requires further studies, including the use of a higher power laser (e.g., the IOTA/FAST laser [4]), use of large bore solenoid with uniform magnetic field and exploration of other types of MCPs – with different materials, dimensions and other arrangements to generate higher pulsed current densities.

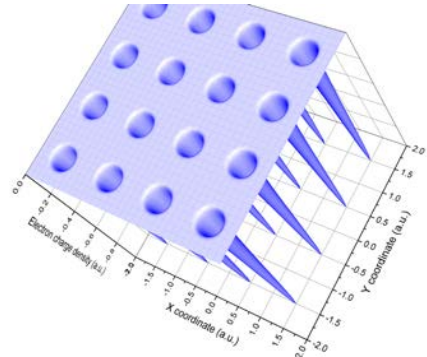

Figure 7: Electron multi-beams configuration [2].

Higher density MCP based electron sources can be used in electron lenses or in the "electron multi-beams" (EMB) – see Fig.7. The EMBs might have many applications in accelerators - for the halo control via modification of beam dynamics of larger amplitude particles; for collimation; for smearing, or homogenization of the accelerator beam phase-space; for the beam center and beam halo diagnostics as the electron beamlet positions after the interaction can be easily measured; and they may be even considered for a kind of sub-coherent beam phase space manipulations. Some of these ideas can be tested at the electron lens setup which is being developed for the IOTA ring at Fermilab [4].